# A DFT based first-principles investigation of the physical properties of $Bi_2Te_2Se$ topological insulator

Md. Asif Afzal, S. H. Naqib*

Department of Physics, University of Rajshahi, Rajshahi 6205, Bangladesh

*Corresponding author e-mail: salehnaqib@yahoo.com

**Abstract**

A topological insulator possesses a bulk energy gap splitting the lowest empty band from the highest occupied electronic band. The electronic states at the surface (or edge in two dimensions), on the other hand, of a topological insulator are gapless and are protected by the time reversal symmetry. Such systems are promising for variety of optoelectronic, superconducting, thermoelectric and quantum computation related applications. We have studied elastic, mechanical, electronic, optical properties, bonding character and the electronic charge density distribution of ternary $Bi_2Te_2Se$ topological insulator. The compound under study is mechanically stable and elastically anisotropic. The electronic band structure calculations reveal high degree of anisotropy in the energy dispersion. Electronic effective mass is high in the *c*-direction compared to that in the *ab*-plane. The optical constants show moderate level of variation with respect to the polarization of the electric field of the incident radiation. The optical spectra are consistent with the electronic band structure and electronic density of states features. Both electronic band structure and optical constants show clear indications of a direct band gap of 0.610 eV for $Bi_2Te_2Se$. It is also found that $Bi_2Te_2Se$ possesses high refractive index at low photon energies in the infrared and visible region. It has low reflectivity in the ultraviolet region. $Bi_2Te_2Se$ absorbs photons strongly in the ultraviolet energies. All these features make $Bi_2Te_2Se$ suitable for diverse class of optoelectronic device applications.

# 1. Introduction

The discovery of distinctive phases of matter is a persisting theme in condensed matter physics. The topological order in solids was found with the study of the quantum Hall effect over 30 years ago [1] but the latest quest for the different topological materials started with the discovery of topological insulators (TIs) [2, 3]. TI is a distinct phase of a material which has a bulk band gap, but has gapless electronic conducting states at the edge or surface of the material which is protected by the time reversal symmetry [4]. These conducting states are due to the combination of spin orbit interaction and time reversal symmetry [5]. These exciting features led to the discovery of different types of topological materials, such as, topological semimetals and topological superconductors [6, 7].

The first 3D TI was the semiconducting alloy $Bi_{1-x}Sb_x$ was found by both theoretical and experimental studies [8], which had complicated surface electronic structure and narrow band gap. These limitations started the search for new TIs. The second generation 3D TI materials were found to be: $Bi_2Te_3$, $Bi_2Se_3$, $Sb_2Te_3$ and $Bi_2Te_2Se$. These materials have topological band structure. A second generation 3D topological material is $Bi_2Te_3$, which offers potential for the topologically protected behavior in ordinary crystals at room temperature and zero magnetic field [9]. The materials $Bi_2Te_3$, $Bi_2Se_3$, and $Sb_2Te_3$ are known as binary compounds. These binary compounds are known for their great thermal properties. They have excellent thermoelectric performance due to the narrow band gap electronic structure and low lattice thermal conductivity [10 – 13]. $Bi_2Te_2Se$ is ternary compound, which has very similar characteristics as the binary $Bi_2Te_3$ compound. It has very robust topological nature of electronic states at the surface with bulk Dirac cone features and is a very promising material for the applications in nanoelectronics and spintronics [14].

The 3D TI materials $Bi_2Te_3$, $Bi_2Se_3$, $Sb_2Te_3$ have been studied extensively compared to the ternary $Bi_2Te_2Se$ compound. The binary compounds are studied via the first principles methods [15 – 17]. These investigations explored the mechanical, electronic and optical properties of binary materials. The $Bi_2Te_2Se$ compound is a new material compared to the other TIs. It has been studied experimentally and theoretically to a limited extent. The experimental studies found bulk resistivity and quantum oscillations in 2010 [18]. The doped compound can be very efficient in surface state transport and fabrication of the surface-state-based electronic devices which was found experimentally in 2011 [19]. The robustness of the surface states and band structure of the compound was studied in 2012 [14]. The optical properties of the Bi-based materials were studied experimentally in 2012 [20]. The first principles study showed the topological behavior and very high thermoelectric performance of $Bi_2Te_2Se$ compound in 2015 [21].

As far as our knowledge is concerned, the mechanical and electronic properties of $Bi_2Te_2Se$ have not been studied in details yet, neither theoretically nor experimentally. The optical properties have been studied experimentally but no theoretical study exists. These unexplored properties are very important for the understanding of the material for its possible applications. Besides, $Bi_2Te_2Se$ compound is a layered ternary compound, this type of materials possess many exciting features [22 – 30] for which this particular compound demands attention and needs to be studied in detail.

In this paper we have done a detailed investigation of the mechanical, electronic and optical properties of $Bi_2Te_2Se$ compound. The mechanical properties include the structural and elastic properties. These properties describe the bonding characteristics and it is important for understanding the defect dynamics and failure modes in the compound. The electronic properties include the band structure and electronic density of states which describe the electronic behavior of the material. The electronic density of states shows a possibility of electronic phase transition due to bonding and anti-bonding peaks close to the Fermi level. The optical properties are very important for a material to be used in the optoelectronic device industry.

The rest of the paper is arranged in the following manner: In Section 2, we have briefly described about the computational methodology used in this study. In Section 3, we have described the crystal structure of the compound and presented the results and analysis of all the physical properties we have studied. Finally in Section 4, we have discussed the key findings of the study and summarized the main conclusions of the present study.

## 2. Computational methodology

The most popular practical approach to *ab-initio* modeling of structural and electronic properties of solids is the density functional theory (DFT). Here the ground state of the crystalline system is found by solving the Kohn-Sham equation [31]. The present investigation of the $Bi_2Te_2Se$ system is based on the first principles method. This is based on the DFT with periodic boundary conditions. By solving the Kohn-Sham equation, the ground state energy of the crystalline system was found. Our DFT based calculations were carried out by the quantum mechanical computational prescription implemented in the CAmbridge Serial Total Energy Package (CASTEP) [32]. We have used the generalized gradient approximation (GGA) for electron exchange correlation in CASTEP. This approximation relaxes the lattice constants because of the repulsive core-valance electron exchange correlation. The GGA was chosen in the scheme of Perdew-Burke-Ernzerhof (PBE) [33]. The interaction between the ion and electron were represented by the Vanderbilt-type ultra-soft pseudopotentials. This pseudopotential saves the computational time significantly without compromising the accuracy of the calculations [34]. We have used the Broyden Fletcher Goldfarb Shanno (BFGS) optimization technique [35] to find out the ground state crystal structure of $Bi_2Te_2Se$ TI.

The electronic orbitals used for Bi, Te, and Se atoms to derive the valance and conduction bands are: $Bi[6s^2\ 6p^3]$, $Te[5s^2\ 5p^4]$ and $Se[4s^2\ 4p^4]$. Periodic boundary conditions were used to determine the total energies of each cell. Plane wave basis was used to expand the trial wave functions. The calculations used a plane-wave cutoff energy of 400 eV for the structure optimization. For the sampling of the Brillouin zone (BZ), k-points grids were generated according to the Monkhorst-Pack scheme [36]. The convergence criteria for the structure optimization and energy calculations were set to ultra-fine quality with the k-point mesh of size 7×7×2. Geometry optimization was achieved using convergence thresholds of $5\times10^{-7}$ eV/atom for the total energy, 0.01 eV/Å for the maximum force, 0.02 GPa for maximum stress, and $0.5\times10^{-3}$Å for the maximum displacement. The independent single crystal elastic constants $C_{ij}$, bulk modulus $B$, and shear modulus $G$ were directly calculated by the CASTEP code using the 'stress-strain' method. From the optimized geometry of $Bi_2Te_2Se$, the electronic band structure features were calculated. We have considered the photon induced electronic transition probabilities between different electronic orbitals to obtain all the optical constants. The optical parameters can be extracted from the knowledge of the complex dielectric function, $\varepsilon(\omega) = \varepsilon_1(\omega) + i\varepsilon_2(\omega)$. The real part $\varepsilon_1(\omega)$ of dielectric function $\varepsilon(\omega)$ can be found from the corresponding imaginary part $\varepsilon_2(\omega)$, which directly follows from the Kramers–Kronig relationships.

The imaginary part, $\varepsilon_2(\omega)$, of the dielectric function is calculated within the momentum representation of matrix elements between occupied and unoccupied electronic states by employing the CASTEP supported formula expressed as [24]:

$$\varepsilon_2(\omega) = \frac{2e^2\pi}{\Omega\varepsilon_0} \sum_{k,v,c} |<\psi_k^c|\hat{u}.\vec{r}|\psi_k^v>|^2 \delta(E_k^c - E_k^v - E) \qquad (1)$$

In the above expression, $\Omega$ is the volume of the unit cell, $\omega$ frequency of the incident electromagnetic wave (photon), $e$ is the electronic charge, $\psi_k^c$ and $\psi_k^v$ are the conduction and valence band wave functions at a given wave-vector $k$, respectively. The conservation of energy and momentum during the optical transition is ensured by the delta function. Once the dielectric function $\varepsilon(\omega)$ is known, all the optical parameters such as the refractive index $n(\omega)$, extinction coefficient $k(\omega)$, reflectivity $R(\omega)$, absorption coefficient $\alpha(\omega)$, optical conductivity and loss function $L(\omega)$ can be computed from it. The Debye temperature was calculated from the elastic constants and elastic moduli of $Bi_2Te_2Se$ via the average sound velocity in the solid.

As far as to the best of our knowledge, this study is the first time where various physical properties of $Bi_2Te_2Se$ compound have been studied in detail. In this study, we have not included the spin-orbit coupling (SOC) because SOC mainly affects the surface electronic state related properties of TIs. SOC, on the other hand, only has a minimal effect on various bulk physical properties, such as the optimized crystal structure, elastic constants, mechanical anisotropy, chemical bonding, thermo-physical behavior, bulk optical properties, bulk electronic dispersions etc. [22, 23]. Many different prior studies on different materials with diverse electronic ground states support this supposition [22 – 24, 37 – 41].

### 3. Results and analysis

**a) Structural and elastic properties**

The crystal structure of $Bi_2Te_2Se$ compound is illustrated in Fig. 1. It is a ternary tetradymite compound with trigonal crystal structure. The space group of $Bi_2Te_2Se$ is $R\bar{3}m$ (no. 166). The unit cell consists of 15 (6 Bi, 6 Se and 3 Te) atoms in total. For our calculations, we have used experimental lattice parameters and atomic positions [42]. The atoms are located at different positions such as, Bi position: $6c$(0, 0, 0.3968), Se position: $3a$(0, 0, 0) and Te position: $6c$(0, 0, 0.2116).

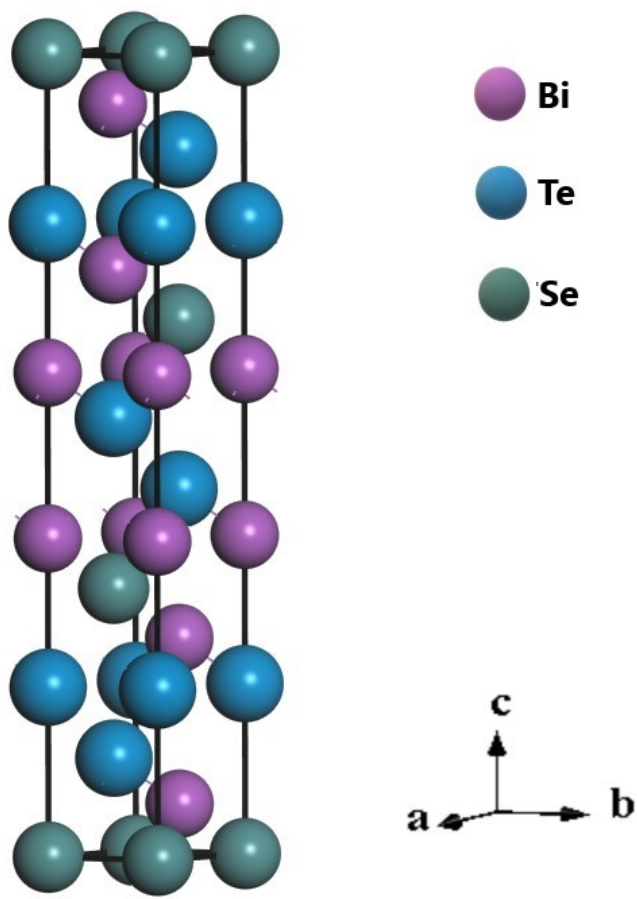

**Figure 1.** The crystal structure of $Bi_2Te_2Se$ compound.

Table 1 shows the optimized lattice parameters along with the available theoretical and experimental results. From the results, we can see that the calculated values show good agreement with the experimental values with a variation of around 3-4%. The theoretical volume of the structural unit is greater than the experimental one because of the GGA, which tends to overestimates the lattice parameters slightly by softening (delocalizing) the electronic orbitals. On the other hand, local density approximation (LDA) tends to underestimate the lattice parameters.

**Table 1:** Optimized lattice parameters of $Bi_2Te_2Se$.

| Nature of study | $a(Å)$ | $c(Å)$ | $c/a$ | $V(Å^3)$ | Ref. |
|---|---|---|---|---|---|
| Theoretical | 4.2952 | 31.7319 | 7.387 | 506.987 | This |
| Experimental | 4.3067 | 30.0509 | 6.977 | 483.243* | [42] |
| Theoretical | 4.279 | 29.910 | 6.989 | 474.276 | [43] |
| Experimental | 4.280 | 29.860 | 6.976 | 473.705 | [43] |

*Calculated from $V = a^2 c \sin(60°)$

$Bi_2Te_2Se$ has trigonal crystal structure. Hence, there are six independent elastic constants ($C_{ij}$) for this particular structure. The elastic constants are: $C_{11}$, $C_{33}$, $C_{44}$, $C_{12}$, $C_{13}$, and $C_{14}$. These elastic constants provide us with important information regarding the mechanical and dynamical behavior of the crystal under stress and the strength of the material under different external conditions. The values of the different elastic constants are shown in Table 2.

Among the six independent elastic constants, $C_{11}$ and $C_{33}$ quantify the response of the crystalline solid under uniaxial strain. $C_{11}$ and $C_{33}$ measure the resistance of a crystal when the uniaxial stresses are along the crystallographic *a* and *c* directions, respectively. For $Bi_2Te_2Se$, $C_{11}$ is greater than $C_{33}$, which suggests that the crystal is more stiff along the *a* and *b* directions compared to the *c* direction. So, the chemical bonding in the *ab*-plane is predicted to be stronger than the bonding in the out-of-plane directions. The elastic constant, $C_{44}$ is related to the shear dominated responses. This is very useful controlling the failure modes in solids. $C_{44}$ also signifies the indentation hardness of materials. The small value of $C_{44}$ indicates the material's inability of resisting the shear deformation in the (100) plane. The other elastic constants $C_{12}$, $C_{13}$, and $C_{14}$ are called the off-diagonal shear components, which are linked with compound's resistance due to various shape distortions. The lowest value of $C_{14}$ comes from the large difference between the values of $C_{11}$ and $C_{44}$. The state of mechanical stability of a structure can be studied by applying the Born-Huang [44] criterion. For trigonal structure, the criteria are [44, 45]:

$$C_{11} - C_{12} > 0;$$

$$(C_{11} - C_{12})C_{44} - 2C_{14}^2 > 0;$$

$$(C_{11} + C_{12})C_{33} - 2C_{13}^2 > 0 \qquad (2)$$

We see that all the stability criteria are satisfied by the elastic constants of Bi$_2$Te$_2$Se, suggesting that the compound is mechanically stable. It is interesting to note that, small negative value of $C_{14}$ has no bearing on the mechanical stability of the TI under study.

**Table 2:** Single crystal elastic constants, $C_{ij}$, of Bi$_2$Te$_2$Se (all in GPa).

| $C_{11}$ | $C_{33}$ | $C_{44}$ | $C_{12}$ | $C_{13}$ | $C_{14}$ | Ref. |
|---|---|---|---|---|---|---|
| 74.2 | 23.7 | 24.9 | 16.2 | 17.2 | -10.6 | This |

The different types of elastic moduli and Poisson's ratio can be calculated from the values of the single crystal elastic constants, $C_{ij}$ [37]. The calculated values of the polycrystalline bulk modulus ($B$), shear modulus ($G$), Pugh's ratio ($B/G$), Young's modulus ($E$), and Poisson's ratio ($\eta$) are displayed in Table 3. The bulk modulus, $B$ and shear modulus, $G$ are obtained from two main approximations, namely, the Voigt (V) [46] and the Reuss (R) [47] schema. The Voigt approximation takes consideration of a continuous strain distribution and a discontinuous stress distribution, which gives us the upper bound of the polycrystalline elastic moduli. On the other hand, The Reuss approximation takes consideration of a continuous stress distribution and a discontinuous strain distribution, which gives us the lower bound of the polycrystalline elastic moduli. The real situation is approximated by the Hill's scheme, which uses the arithmetic average of these two limits and gives proper values suitable for the polycrystalline elastic moduli by considering the right energy considerations [48]. The shear modulus is the resistance of a material to shear or torsional stress, which involves the change of shape without changing the volume. On the other hand, the bulk modulus is resistance of a material to hydrostatic or isotropic stress, which involves the change in volume without changing the shape. So, the shear modulus is a better indicator of the strength of a compound than the bulk modulus. We can see from the above table that, $B$ is greater than $G$. Therefore, the shear component should control the mechanical failure in Bi$_2$Te$_2$Se. The Pugh's ratio is defined by the simple relationship between bulk modulus and shear modulus given by, $B/G$ [49]. This ratio is measure of the ductile or brittle nature of a compound. A high value of this ratio indicates ductility and a low value indicates brittleness of the compound. The critical value of the Pugh's ratio is 1.75. If a material has a value higher than this, it will show ductile behavior and if the value is lower than this then it will exhibit brittleness. From the above table we can see that, for Bi$_2$Te$_2$Se the value of this ratio 1.293. Therefore, Bi$_2$Te$_2$Se is expected to be brittle in nature. The Young' modulus ($E$) is the measure of a material's resistance against tension or compression along its length. It can predict the extension and compression of a material under different tensile stresses. From the above table, we can see that the value of Young's modulus ($E$) is small. So, Bi$_2$Te$_2$Se is weak to resist large tensile stress. The Poisson's ratio ($\eta$) is measure of a material's expansion or compression along the perpendicular direction of the applied stress. It plays a very important role determining the various mechanical properties of crystalline solids. Poisson' ratio is related to the nature of interatomic forces in solids [50]. For solids with the value of $\eta$ in the range of 0.25 to 0.50, the central force interaction dominates. On the other hand, for solids outside this range will have non-central force dominated atomic bondings. Thus, Bi$_2$Te$_2$Se should be a non-central force dominated solid in regards to atomic bonding. $\eta$ can also predict the failure mode of crystalline solids [51]. It has a critical value of 0.26. A value higher than this indicates ductile feature and a lower value indicates brittle nature. So, Bi$_2$Te$_2$Se should have brittle nature according to the value of $\eta$ as seen in Table 3. This is consistent with the result predicted by the Pugh's ratio. The value of $\eta$ for completely metallic bonded compound is around 0.33 and in purely covalent crystal, it is around 0.10. The compound Bi$_2$Te$_2$Se has the Poisson's ratio $\eta = 0.192$. Therefore, it should be dominated by covalent bonding. The low value of $\eta$ predicts the stability against shear. In the absence of prior studies, we cannot compare these values with any previous result.

**Table 3:** Elastic moduli (all in GPa), Pugh's ratio, and Poisson's ratio of Bi$_2$Te$_2$Se.

| $B_V$ | $B_R$ | $B$ | $G_V$ | $G_R$ | $G$ | $B/G$ | $E$ | $\eta$ | Ref. |
|---|---|---|---|---|---|---|---|---|---|
| 30.360 | 22.483 | 26.421 | 23.867 | 16.996 | 20.432 | 1.293 | 48.734 | 0.192 | This |

The elastic anisotropy is a very important feature for crystalline solids. It is used in material design and engineering sciences because of its correlation with the possibility of creation and propagation of microcracks inside the crystals [22, 24] and to predict its mechanical response under variety of external stresses. We have calculated the Zener anisotropy factor $A$, Chung and Buessem anisotropy factor $A^C$, and the shear anisotropy factors: $A_1$, $A_2$, $A_3$. We have also calculated the universal anisotropic index, $A^U$ and d$_E$ using the anisotropy indices of bulk and shear moduli $A_B$ and $A_G$ respectively. All the calculated values are presented in the Table 4. The following relations are used to calculate the anisotropy indices [52, 53]:

$$A = \frac{2C_{44}}{C_{11} - C_{12}}$$

$$A^c = \frac{G_V - G_R}{2G_H}$$

$$A_1 = \frac{4C_{44}}{C_{11} + C_{33} - 2C_{13}}$$

$$A_2 = \frac{5C_{55}}{C_{22} + C_{33} - 2C_{23}}$$

$$A_3 = \frac{4C_{66}}{C_{11} + C_{22} - 2C_{12}}$$

$$A_B = \frac{B_V - B_R}{B_V + B_R}$$

$$A_G = \frac{G_V - G_R}{G_V + G_R}$$

$$A^U = \frac{B_V}{B_R} + 5\frac{G_V}{B_R} - 6 \geq 0$$

$$d_E = \sqrt{A^U + 6}$$

From the above table, we can see that, the values of elastic anisotropy indices are significantly different from unity. So, the studied compound $Bi_2Te_2Se$ is predicted to show highly anisotropic elastic and mechanical characteristics.

A useful physical parameter, the Cauchy pressure, is defined as the difference between two particular elastic constants ($C_{12} - C_{44}$), which gives us an insight into chemical bonding and elastic response of solids [54]. It can indicate both the failure mode and the nature of chemical bonding. If the value of the Cauchy pressure is positive, the material should be ductile and have predominantly metallic bonding and if the value is negative then it should be brittle and have significant covalent bonding. In the studied compound, the Cauchy pressure is negative. The material should be brittle and should have prominent covalent bonding.

All the above elastic measures regarding ductility and brittleness for this material are consistent with each other and showed that, Bi$_2$Te$_2$Se crystal should be brittle, have covalent bonding and it is a non-central force dominated material.

**Table 4:** Calculated indices of elastic anisotropy of Bi$_2$Te$_2$Se.

| $A$ | $A^C$ | $A^L$ | $A_1$ | $A_2$ | $A_3$ | $A_B$ | $A_G$ | $A^U$ | $d_E$ | Ref. |
|---|---|---|---|---|---|---|---|---|---|---|
| 0.86 | 0.168 | 0.816 | 1.569 | 1.569 | 0.998 | 0.149 | 0.168 | 2.39 | 2.896 | This |

The bulk moduli for the single crystal along different crystallographic axes [55] have been evaluated and are shown in Table 5. $B_{\text{relax}}$ is the single crystal isotropic bulk modulus, which has the same value we found from the Reuss approximation. $\alpha$ and $\beta$ can be characterized as the relative change of $b$ and $c$ axis as a function of the deformation of the $a$ axis. The linear bulk modulus along the crystallographic axes can also be obtained from the pressure gradient. The values of $B_a$ and $B_b$ are same and high compared with the value of $B_c$, which suggests that the compound is much stiffer in the $a$ and $b$ directions and it is more compressible along $c$ direction. The values also suggest that, there is a significant amount of bonding anisotropy in the compound and the compound is more anisotropic in the $c$-direction rather than within the $ab$-plane. The machinability index can be defined as the ratio between bulk modulus to $C_{44}$ [56]. This parameter determines the use of the compound in the materials science and engineering field with different shapes. Furthermore, the hardness of a solid can be expressed as [53]:

$$H_V = 2\,(k^2 G)^{0.585} - 3$$

The calculated value of the hardness for Bi$_2$Te$_2$Se is 5.647 GPa. The value indicates the material is moderately hard in nature.

**Table 5:** The bulk modulus ($B_{\text{relax}}$ in GPa), bulk modulus along the crystallographic axes $a$, $b$, $c$ ($B_a$, $B_b$, $B_c$), $\alpha$, $\beta$ and the machinability index ($\mu_M$) of Bi$_2$Te$_2$Se.

| $B_{\text{relax}}$ | $B_a$ | $B_b$ | $B_c$ | $\alpha$ | $\beta$ | $\mu_M$ |
|---|---|---|---|---|---|---|
| 22.483 | 237.286 | 237.286 | 27.739 | 1.0 | 8.554 | 1.07 |

## b) Debye temperature and thermal parameters

The Debye temperature, $\theta_D$ is a fundamental characteristic parameter of solids. It is the highest equivalent temperature to the angular frequency, which can be sustained by a single normal mode of lattice vibration. This fundamental parameter is associated with many other important physical properties, such as phonon heat capacity, bonding strengths, phonon thermal conductivity, vacancy formation energy, melting temperature, etc. In a crystal, the excitations at low temperatures arise only from acoustic vibrations. At higher temperatures, the optical modes are excited. The Debye temperature, $\theta_D$ is calculated from the mean sound velocity, $v_m$ and mass density, $\rho$ of the solid. The calculated elastic constants were also used to find the $\theta_D$. We have used the following expressions [57] to estimate $\theta_D$:

$$\theta_D = \frac{h}{k_B}\left[\left(\frac{3n}{4\pi}\right)\frac{N_A \rho}{M}\right]^{\frac{1}{3}} v_m$$

where, $h$ denotes Planck's constant, $k_B$ denotes Boltzmann's constant, $N_A$ denotes Avogadro's number, $\rho$ denotes mass density, $M$ denotes the molecular weight and $n$ denotes the number of atoms in the cell.

The mean sound velocity, $v_m$ in the crystal is calculated from,

$$v_m = \left[\frac{1}{3}\left(\frac{1}{v_l^3} + \frac{1}{v_t^3}\right)\right]^{\frac{-1}{3}}$$

where, $v_l$ and $v_t$ represent the longitudinal and transverse modes of sound velocities. These can be calculated from the bulk modulus, $B$ and shear modulus, $G$:

$$v_l = \left[\frac{3B + 4G}{3\rho}\right]^{\frac{1}{2}}$$

and

$$v_t = \left[\frac{G}{\rho}\right]^{\frac{1}{2}}$$

An empirical formula for calculating the melting temperature of different crystals was proposed by Fine *et al.* [58] using the elastic constants of the crystal:

$$T_m = 354 + 1.5(2C_{11} + C_{33})$$

The behavior of atoms inside a crystal can be determined with the help of thermal conductivity when the crystal is heated or cooled. The hypothetical lowest value of inherent thermal conductivity is defined by the minimum thermal conductivity of the crystal. It can be evaluated by using the average sound velocity and the formula [59] expressed as:

$$\kappa_{min} = k_B v_m \left(\frac{nN_A\rho}{M}\right)^{\frac{2}{3}}$$

**Table 6:** Calculated mass density ($\rho$ in gm cm$^{-3}$), longitudinal, transverse and average sound velocities ($v_l$, $v_t$, and $v_m$, all in km s$^{-1}$), Debye temperature ($\theta_D$ in K), melting temperature ($T_m$ in K) and minimum thermal conductivity ($\kappa_{min}$ in Wm$^{-1}$K$^{-1}$) of Bi$_2$Te$_2$Se.

| $\rho$ | $v_l$ | $v_t$ | $v_m$ | $\theta_D$ | $T_m$ | $\kappa_{min}$ |
|---|---|---|---|---|---|---|
| 7.39 | 2.694 | 1.667 | 1.832 | 168.442 | 612.12 | 0.2415 |

The calculated Debye temperature ($\theta_D$), melting temperature ($T_m$) and minimum thermal conductivity ($\kappa_{min}$) of Bi$_2$Te$_2$Se are shown in Table 6. A high value of Debye temperature indicates a high phonon thermal conductivity. From Table 6, we can see that Bi$_2$Te$_2$Se has a comparatively low value of Debye temperature. This suggests that, the Bi$_2$Te$_2$Se compound has a flexible lattice and comparatively low phonon thermal conductivity. The melting temperature is also in the lower side, which describes it's limitation to be used at elevated temperatures. The minimum thermal conductivity is also consistent with low Debye temperature which suggests a low phonon thermal conductivity.

### c) Electronic band structure and Density of states

**Electronic band structure**

The calculated electronic band structure of Bi$_2$Te$_2$Se TI along the high-symmetry directions, $\Gamma$-$A$-$H$-$K$-$\Gamma$-$M$-$L$-$H$ in the Brillouin is plotted in Fig. 2. The vertical line denotes the Fermi energy $E_F$, which is set at the energy 0 eV. The energy dispersion curves below the Fermi level constitute the valance band and the curves above the Fermi level are the conduction band of the compound. The band structure calculations show that, at zero pressure, the compound has a direct band gap with an energy gap of 0.610 eV. The direct band gap implies that the top of the valance band is at the same $k$-value as the bottom of the conduction band.

The curves along *A-H*, *H-Γ* and *Γ-M* directions are highly dispersive, which indicates small effective mass of electrons and high mobility of the charge carriers in these bands. The energy dispersion curves along *Γ-A*, *H-K* and *M-L* are almost non-dispersive, which indicates very high effective mass of charge carriers and consequently very low mobility in these directions. This implies a significant anisotropy in charge transport in the in-plane and out of the *ab*-plane directions. So, the layered structure of this compound has led to notable anisotropy in the electronic band structure. There is almost linear energy dispersion at the *Γ*-point in the conduction band close to Fermi level. This is Dirac cone feature appropriate to TIs. Therefore, from the band structure properties we see that the compound under consideration should be a weak topological insulator.

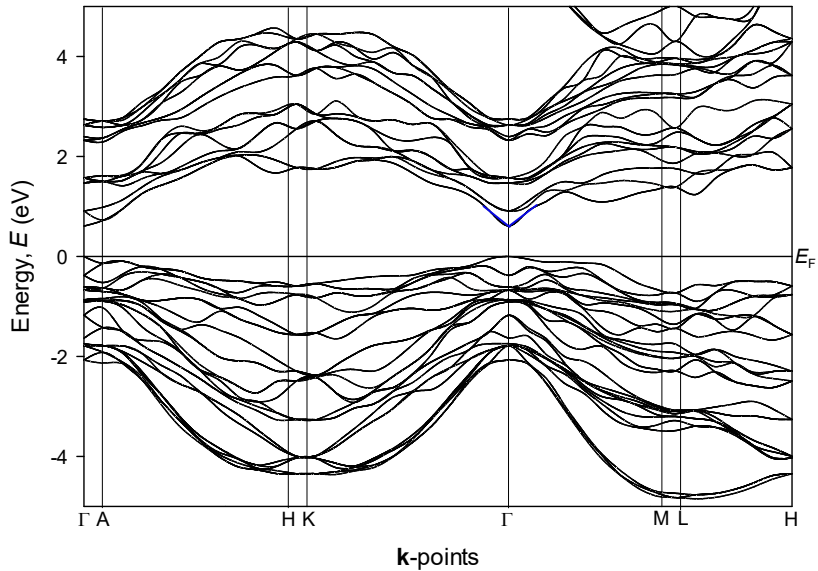

**Figure 2:** Electronic band structure of $Bi_2Te_2Se$ compound along the high symmetry points of the *k*-space within the first Brillouin zone. The Fermi level is located at 0 eV. We have marked a representative part of linear dispersion with blue straight lines near the *Γ*- point.

**Electronic energy density of states**

The calculated total density of states (TDOS) and atom resolved partial density of states (PDOS) of $Bi_2Te_2Se$ system is presented in Fig.3 as a function of energy, $(E - E_F)$. The vertical straight line in the figure represents the Fermi level, $E_F$. The valance band consists of Bi-6*p*, Te-5*p* and Se-4*p* electronic states. These three atomic orbitals all contribute to the valance band with Se-4*p* having the highest contribution. Near the Fermi level, the contribution to the total DOS comes from three different orbitals. They are Bi-6$s$, Te-5$p$ and Se-4$p$, among these the Se-4$p$ has the largest contribution. So, the charge transport and bonding properties are dominated by the hybridization of these aforementioned electronic states.

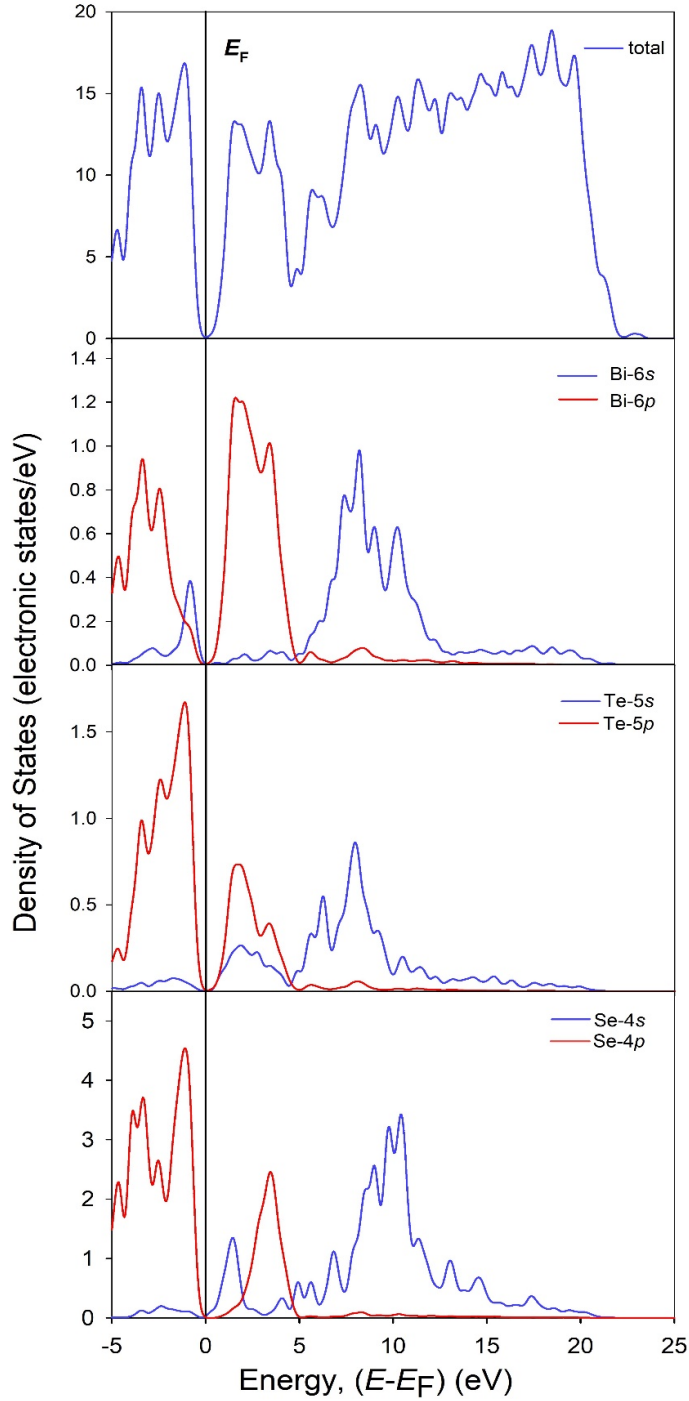

**Figure 3:** TDOS and PDOS of Bi$_2$Te$_2$Se. The vertical line shows the Fermi energy.

The higher energy conduction band in the total DOS above the Fermi energy is formed by the Bi-6$s$, Te-5$s$ and Se-4$s$ electronic states. Here, the major contributors are the Bi-6$s$, Te-5$s$ electronic states. Another important feature of the TDOS which can be seen from Fig. 3 is that the bonding and anti-bonding peaks are quite close to the Fermi level, $E_F$. The bonding peak is at the left of the Fermi level, $E_F$. There is covalent bonding between the atoms of this material that can be understood with the minimum between the bonding and anti-bonding peaks, known as pseudogap, which is very close to the Fermi level. So, by tuning the compound with doping or changing the temperature or pressure, we can change the electronic phase of the compound. So, there is a possibility of electronic phase transition in $Bi_2Te_2Se$ due to external stimulus.

**d) Charge density distribution**

For the study of the electronic charge density distribution within the $Bi_2Te_2Se$ compound, the valance electronic charge density map has been studied and is shown in Fig. 4 projected in the (100) crystallographic plane. The color scale on the right side of the electronic charge density map represents the total electron density. Here blue and red color represents the highest and lowest electron density respectively. The electron charge density results presented in this section are grossly consistent with the Mulliken and Hirshfeld charge analyses. The bonding nature in the $Bi_2Te_2Se$ compound can be understood from the charge density distribution. From Fig. 4 we can see that, the charge distribution around the atoms Bi, Te and Se are not completely spherical. This suggests that covalent bondings are present in the $Bi_2Te_2Se$ compound. There is a huge amount of charge accumulation between the Bi-Te bonding. The electron density is much higher on the Te side, because Te has a higher electronegativity compared to Bi. This is a clear sign of ionic bonding between Te and Bi atoms. This is also consistent with the Mulliken/Hirshfeld charge analysis. Therefore, we predict a strong admixture of covalent and ionic bondings in the $Bi_2Te_2Se$ TI compound.

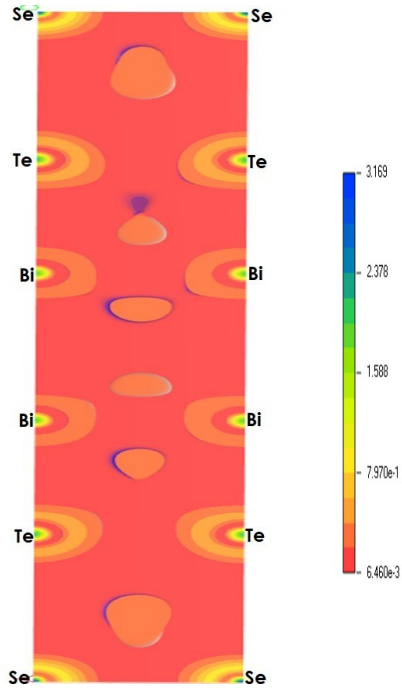

**Figure 4:** The charge density distribution in $Bi_2Te_2Se$ compound projected in the (100) plane.

### e) Bond population analysis

We have analyzed both Mulliken bond population analysis [60] and the Hirshfeld population analysis [61]. The results of these analyses are represented in Table 7. The Mulliken charge densities of Bi, Te and Se in $Bi_2Te_2Se$ compound are 0.34, -0.08 and -0.51 electronic charges, respectively. From these values we can see that, Bi atom give up some electrons to Te and Se atoms. All these values are deviated from the normal value in a purely ionic state of the respective atoms. This deviation is due to the covalent bonding among the atoms in the compound. There is covalent bonding among the Bi, Te and Se atoms because of the hybridization of the Bi-$6p$, Te-$5p$ and Se-$4p$ electronic orbitals. This is predicted by the PDOS features shown in Fig. 3. From Table 7 we can see that, the band spilling parameter has a relatively low value. This reflects the very low value of TDOS at Fermi level. The transfer of electrons between different atoms in the compound also reflects to the partial presence of ionic bonding characteristics.

xvii

**Table 7:** Charge spilling parameter (%), orbital charges (electron), atomic Mulliken charges (electron), effective valance (electron) and Hirshfeld charge (electron) in $Bi_2Te_2Se$ compound.

| Atoms | Charge spilling (%) | s | p | d | Total | Mulliken charge | Hirshfeld charge |
|---|---|---|---|---|---|---|---|
| Bi |  | 2.11 | 2.56 | 0.00 | 4.66 | 0.34 | 0.21 |
| Te | 0.43 | 1.79 | 4.29 | 0.00 | 6.08 | -0.08 | -0.09 |
| Se |  | 1.88 | 4.63 | 0.00 | 6.51 | -0.51 | -0.23 |

Furthermore, Table 8 represents the calculated band overlap populations and bond lengths between the atoms of $Bi_2Te_2Se$. The overlap population has higher negative value in Bi-Te bonding and lower negative value in the Se-Bi bonding. These values suggest that, there is a significant interaction between the atoms and they have anti-bonding nature. Bond length is a measure of the strength of chemical bonds. In general, shorter bond length implies stronger bonding and higher bond hardness. Our calculations predict that, the bonding strengths for Bi-Te bond and Se-Bi bond are quite close and comparable with each other.

**Table 8:** Calculated bond overlap population and bond lengths (Å) for $Bi_2Te_2Se$.

| Bond | Population | Length |
|---|---|---|
| Te 3 -- Bi 5 | -8.22 | 3.04268 |
| Te 5 -- Bi 3 | -8.22 | 3.04268 |
| Te 2 -- Bi 4 | -8.22 | 3.04268 |
| Te 4 -- Bi 2 | -8.22 | 3.04268 |
| Te 1 -- Bi 6 | -8.22 | 3.04268 |
| Te 6 -- Bi 1 | -8.22 | 3.04268 |
| Se 3 -- Bi 2 | -2.98 | 3.07209 |
| Se 2 -- Bi 6 | -2.98 | 3.07209 |
| Se 1 -- Bi 3 | -2.98 | 3.07209 |
| Se 1 -- Bi 5 | -2.98 | 3.07209 |
| Se 3 -- Bi 4 | -2.98 | 3.07209 |
| Se 2 -- Bi 1 | -2.98 | 3.07209 |

**f) Optical properties**

The response of a material to the incident electromagnetic radiation can be analyzed from the frequency dependent optical parameters. The study of the optical properties of solids reveals important facets of band structure, localized charged defects, lattice vibrations and impurity levels. These properties are very important for the investigation of the response of solids at different photon energies. For possible optoelectronic and photovoltaic device applications of solids, information regarding its response to the infrared, visible and ultraviolet spectra is very important. These responses are completely understood by the various energy (or equivalently frequency, $f$) dependent optical constants. These optical constants are dielectric function $\varepsilon(\omega)$, refractive index $n(\omega)$, conductivity $k(\omega)$, reflectivity $R(\omega)$, absorption coefficient $\alpha(\omega)$ and the energy loss function $L(\omega)$, where $\omega = 2\pi f$, is the angular frequency. The different optical constants of $Bi_2Te_2Se$ compound are shown in the Fig. 5 for incident energy up to 30 eV and the electric field polarizations along [100] and [001] directions. We have considered the default settings in CASTEP for the calculation of the optical constants appropriate for semiconductors and insulators. Therefore, we have taken a Drude damping (relaxation energy) of 0.05 eV and a Gaussian smearing of 0.5 eV for the calculations of the optical constants.

Fig. 5 (a) shows the absorption coefficient $\alpha(\omega)$ of $Bi_2Te_2Se$. The absorption coefficient starts at around 0.60 eV, which confirms the semiconducting nature of the compound. The onset of optical absorption agrees completely with band structure calculations. The maximum absorption of the incident radiation occurs around ~8.0eV and falls down at the higher energies around ~ (20 – 25) eV for both the polarizations of the electric field. From the absorption spectra we can also see that this compound can absorb the ultraviolet radiation efficiently. Fig. 5 (b) shows the loss function $L(\omega)$ of $Bi_2Te_2Se$. This particular parameter is a measure of the energy loss of a fast electron moving through the solid. The loss peak is due to the plasma resonance. The loss function peak was found at around ~18 eV. This is the characteristic plasmon energy. The frequency dependent reflectivity is shown in the Fig. 5 (c). There is a high reflectivity band in the energy region encompassing 2.0 eV – 6.0 eV. The maximum and minimum reflectivities are in the ultraviolet region for both the polarizations. In the mid ultraviolet to deep ultraviolet region $Bi_2Te_2Se$ has low reflectivity for both the polarizations; therefore, this material can be used as an anti-reflection coating material. The reflectivity decreases sharply from the energy close to the plasma frequency, consistent with the position of energy loss peak. In the higher energy region above the plasma frequency the material shows transparent optical behavior.

Fig. 5 (d) shows the real and imaginary parts of the calculated optical conductivity. The optical conductivity rises sharply around ~0.60 eV for both the polarizations. This implies the semiconducting nature of the material, which agrees with results found from the electronic band structure calculations and the optical absorption profile. The optical conductivity increases initially with increasing energy above 0.60 eV due to photon induced generation of electron-hole pairs. The refractive index is a hugely important parameter for photonic device applications and also for optical waveguides. Both the real and imaginary parts of the refractive index are shown in Fig. 5 (e). The phase velocity of the electromagnetic wave inside the compound is measured by the real part of the refractive index, the imaginary part, on the other hand, determines the amount of attenuation of electromagnetic wave inside the compound. From Fig. 5 (e) it is noticed that the refractive index has a very high value (~5) at low energies in the infrared and visible regions. This high value suggests that this material might be used in the optimization of the electronic display devices such as in LCD, OLED and QLED. Finally, Fig. 5 (f) represents the real and imaginary parts of the dielectric constants. The real part is associated with the electrical polarization of the material. The imaginary part is associated with the dielectric loss. The dielectric constants show the onset at energy is around ~0.60 eV, which is also found from the calculated electronic band structure and other optical properties. The real part crosses the zero value at the energy around plasma frequency, which is found from the loss peak near 18 eV. At energies higher than this, the real part of the dielectric function goes to unity and the imaginary part goes to very low value. Hence, the material should become transparent for photons with energies above 18 eV.

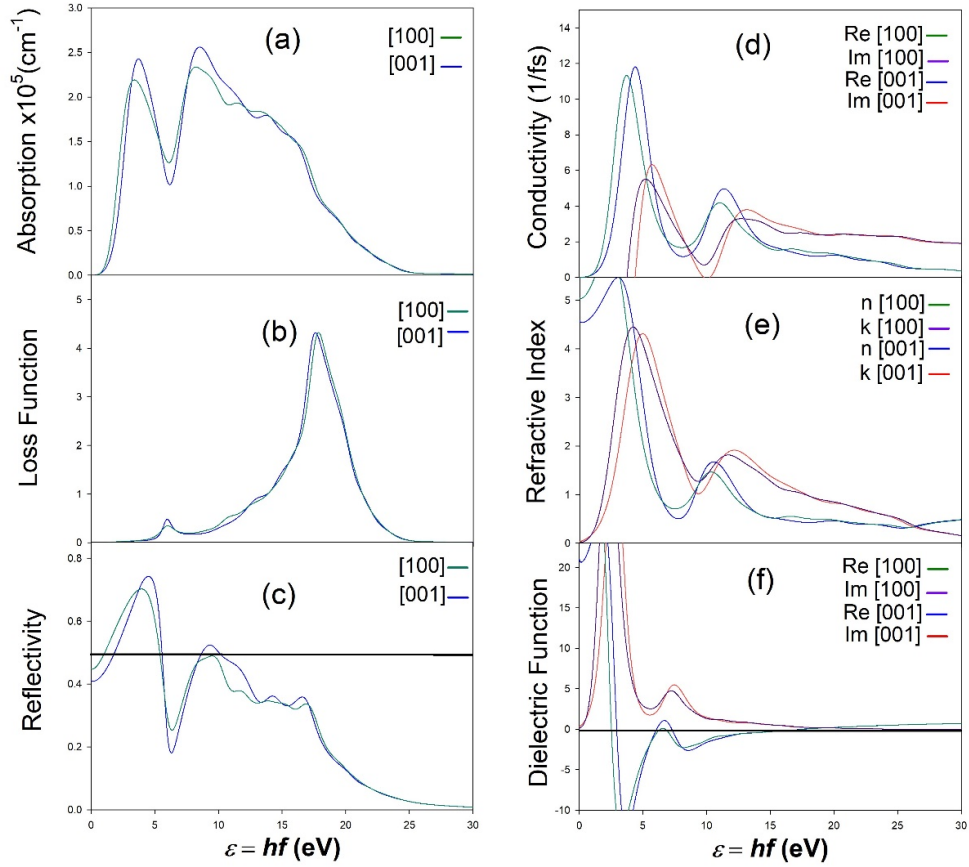

**Figure 5:** The frequency dependent (a) absorption coefficient, (b) loss function, (c) reflectivity, (d) optical conductivity, (e) refractive index and (f) dielectric constant of $Bi_2Te_2Se$ with electric field polarization vectors along [100] and [001] directions.

The spectral features of all the optical parameters of $Bi_2Te_2Se$ are almost independent of the polarization of the incident electromagnetic wave. Therefore, the topological insulator under study is optically almost isotropic.

### 4. Discussion and conclusions

We have investigated the structural, mechanical, electronic, thermodynamical and optical properties of the $Bi_2Te_2Se$ compound by using the first principles DFT based calculations. The calculated structural properties are in a good agreement with the experimental results. The elastic constants confirm the mechanical stability of the compound. The analyses of mechanical properties predict the anisotropic nature of the compound. The brittleness, covalent bonding and non-central force domination of this compound are confirmed from the Poisson's ratio, Pugh's ratio, and Cauchy pressure values. The machinability index of Bi$_2$Te$_2$Se is rather low, especially when compared to many other ternary layered solids [22 – 30]. The low value of Debye temperature indicates low phonon thermal conductivity. The hardness of the material is comparatively low. All these properties suggest that Bi$_2$Te$_2$Se TI is rather a soft compound.

The electronic properties include the electronic band structure and electronic energy density of states. The band structure reveals a direct band gap of 0.610 eV. This low energy gap implies that the material is capable of absorbing significant part of solar radiation. The compound has a significant amount of anisotropy in charge transport in the out of the *ab*-plane. From the band structure calculations, we can see almost linear dispersion at the *Γ*-point in the conduction band close to the Fermi level. This is Dirac cone feature, a signature feature of topological material, previously shown in different studies [13, 22, 23]. The electronic energy density of states discloses that the *p*-orbital dominates in the valance bad and *s*-orbital dominates in the conduction band. So, the charge transport and bonding properties are dominated by the hybridization between these orbitals. The possibility of the electronic phase transition is there, because the Fermi level lies very close to the middle between the bonding and anti-bonding peaks. The charge density distribution suggests covalent bonding in the compound with some ionic contributions; this is also confirmed from the Mulliken and Hirshfeld population analyses.

The optical properties of the compound are studied for two different planes of polarizations [100] and [001] of the electric field. The absorption coefficient spectrum confirms the semiconducting nature of the compound. The reflectivity spectra show that, the material can be used as a good reflector of the visible radiation, due to the maximum reflectivity in the visible region. The conductivity spectra also confirm the semiconducting nature of the material. The refractive index of the material is high in the visible and infrared region. All the optical parameters showed little anisotropy with respect to the polarization planes of the incident electromagnetic radiation. The reflectivity and optical conductivity results obtained in this study exhibit fair agreement with those found experimentally in Ref. [20].

We hope that, the results presented in this study will encourage the scientific community to explore the physical properties of the topological insulator Bi$_2$Te$_2$Se in greater detail, both experimentally and theoretically, so that the full potential of this interesting compound for possible applications can be unlocked.

**Acknowledgements**

S. H. N. acknowledges the research grant (1151/5/52/RU/Science-07/19-20) from the Faculty of Science, University of Rajshahi, Bangladesh, which partly supported this work.

**Data availability**

The data sets generated and/or analyzed in this study are available from the corresponding author on reasonable request.

**Author Contributions**

**Additional Information**

**Competing Interests**